\documentclass[aps,prd,onecolumn,nofootinbib,notitlepage]{revtex4-1}
\usepackage{amsmath,amssymb,amstext,graphicx,color,xfrac,braket,accents,tabulary,booktabs,xcolor,bbold}
\usepackage[T1]{fontenc}
\begin{document}
\baselineskip=13pt
\hfill CALT-TH-2020-40
\hfill

\vspace{0.5cm}
\thispagestyle{empty}

\newcommand{\be}{\begin{equation}}
\newcommand{\ee}{\end{equation}}
\newcommand{\bea}{\begin{eqnarray}}
\newcommand{\eea}{\end{eqnarray}}
\newcommand{\hst}{\widetilde{\mathcal{H}}} 
\newcommand{\iso}{\dot{=}}
\newcommand{\Dim}{\mathrm{dim\,}}
\newcommand{\Tr}{\mathrm{Tr}}
\newcommand{\hs}{\mathcal{H}} 
\newcommand{\ham}{\hat{H}}
\newcommand{\intham}{\widehat{H}_{\rm{int}}}
\newcommand{\selfham}{\widehat{H}_{\rm{self}}}
\def\bra#1{\langle #1\rvert}
\def\ket#1{\lvert #1\rangle}
\newcommand{\draftnote}[1]{\textbf{\color{red}[#1]}}
\newcommand{\psiuu}{\psi_{\uparrow\uparrow}}
\newcommand{\psiud}{\psi_{\uparrow\downarrow}}
\newcommand{\psidu}{\psi_{\downarrow\uparrow}}
\newcommand{\psidd}{\psi_{\downarrow\downarrow}}

\title{Energy Non-Conservation in Quantum Mechanics}
\author{Sean M. Carroll$^1$ and Jackie Lodman$^2$}
\affiliation{$^1$Walter Burke Institute for Theoretical Physics, California Institute of Technology, 
and Santa Fe Institute; \rm{seancarroll@gmail.com}\\
$^2$\it{California Institute of Technology, and Department of Physics, Harvard University}; \rm{jlodman@g.harvard.edu}}

\begin{abstract}
We study the conservation of energy, or lack thereof, when measurements are performed in quantum mechanics.
The expectation value of the Hamiltonian of a system changes when wave functions collapse in accordance with the standard textbook (Copenhagen) treatment of quantum measurement, but one might imagine that the change in energy is compensated by the measuring apparatus or environment.
We show that this is not true; the change in the energy of a state after measurement can be arbitrarily large, independent of the physical measurement process.
In Everettian quantum theory, while the expectation value of the Hamiltonian is conserved for the wave function of the universe (including all the branches), it is not constant within individual worlds.
It should therefore be possible to experimentally measure violations of conservation of energy, and we suggest an experimental protocol for doing so.
\end{abstract}

\maketitle

\section{Introduction}

Quantum mechanics is harder to make sense of than most other physical theories, primarily because of the measurement problem \cite{Wallace2007-WALTQM}.
For this reason among others, it is useful to think carefully about what actually happens when quantum measurements occur.
An important aspect of this question is the status of energy conservation.

There is more than one way one could interpret ``energy conservation'' in a quantum context.
Consider an isolated system with a time-independent Hamiltonian $\ham$.
We can distinguish at least three possibilities.
First, one could take the attitude that energy is only {defined} when the system is in an energy eigenstate, where $\ham\ket{\psi} = E\ket{\psi}$.
If the system under consideration is undisturbed, its energy will be constant.
But this is because there are no dynamics at all, as the state obeys $\ket{\psi(t)} = e^{-iE t }\ket{\psi(0)}$.
(Unless explicitly specified otherwise, henceforth we will set $\hbar=c=1$.)
The system stays (modulo an overall phase) in whatever state it started in; energy doesn't change because nothing changes.
So this is a notion of energy conservation, but a somewhat trivial one, applicable only to highly non-generic and dynamically uninteresting states.

Another possibility is to imagine that there is a true and conserved energy of the system, but that we cannot deduce what it is from the wave function $\ket{\psi}$, perhaps because there are hidden variables or because of a fundamentally epistemic notion of the wave function.
In that case it is conceivable that energy is precisely conserved, but only in a not-very-useful sense, since we can't know what it is.

A third option suggests itself if the wave function is a complete representation of the physical system, as in ontic approaches such as Everettian (Many-Worlds) quantum mechanics or objective-collapse models.
There it makes sense to associate an energy to any state $|\psi\rangle$ by taking the expectation value of the Hamiltonian,
\be
  E_\psi \equiv \langle \psi|\ham |\psi\rangle.
  \label{energydef}
\ee
This quantity is manifestly conserved under unitary evolution by a time-independent Hamiltonian, 
\be
|\psi(t)\rangle = \exp(-i\ham t)|\psi(0)\rangle.
\ee
This could equivalently be characterized as a consequence of Noether's Theorem, or as the fact that the Hamiltonian commutes with itself.

But measurement and wave function collapse are non-unitary processes, not bound by Schr\"odinger evolution.
At first glance, there is no reason to expect that $E_\psi$ is conserved in such circumstances, and indeed this quantity changes in magnitude according to the textbook quantum recipe.
Consider a system with Hamiltonian $\ham$ and energy eigenstates $|E_n\rangle$ with eigenvalues $E_n$, and imagine that (for some reason) we know that the system is in a state
\be
  |\psi_1\rangle = \alpha |E_1\rangle + \beta |E_2\rangle,
  \label{energysuperposition}
\ee
with $|\alpha|^2+|\beta|^2 = 1$ and $E_1\neq E_2$.
Now perform a measurement in the energy eigenbasis.
The wave function will collapse to $|\psi_2\rangle$ with probabilities given by the respective amplitudes:
\begin{align}
  |\psi_2\rangle = |E_1\rangle\,, \  \mathrm{probability\ }|\alpha|^2;\qquad \qquad 
  |\psi_2\rangle = |E_2\rangle\,,\  \mathrm{probability\ }|\beta|^2.
\end{align}
In these cases, the energy (\ref{energydef}) changes from $|\alpha|^2 E_1 + |\beta|^2 E_2$ to either $E_1$ or $E_2$, respectively. 
In that sense, energy is not conserved.
The effect need not be small: we might consider the two (approximate) energy eigenstates to represent a bowling ball at rest and one moving at high velocity, associated with very different energies, while the measurement could be carried out by a single photon.

Nevertheless, energy seems to be conserved in our experience, at least to very high precision. 
Why don't we see violation of energy conservation all the time? 
An obvious possibility is that the above analysis neglects any energy transferred to the measurement apparatus or environment. 
One purpose of this paper is to show that this is not the case, and that the total energy of the system plus apparatus plus environment can change.
Alternatively, one might not think there is any reason energy should be conserved in quantum measurements, especially in epistemic approaches.
But regardless of one's intuition, the expectation value of the Hamiltonian is a well-defined quantity in any quantum state that is exactly conserved under unitary evolution.
Its non-conservation is associated strictly with measurement and wave function collapse; furthermore, it is approximately conserved in the real world, a feature that seems puzzling if the quantity is meaningless.
Given the unsettled status of these processes, we feel it is worth working to better understand what happens to energy when quantum states are measured.

There has been relatively little discussion of this phenomenon in the literature, although it has been noticed.
Hartle et al.~\cite{hartle1995conservation}, in part based on an argument from Griffiths \cite{griffiths}, argued for a sense of energy conservation for closed quantum systems in the decoherent-histories formalism.
They showed that histories could not decohere if an energy measurement (perhaps with a range of allowed values) at one time disagreed with another measurement at another time.
Pearle \cite{Pearle:2000qb} notes explicitly that ``the collapse postulate of standard quantum theory can violate conservation of energy-momentum,'' and investigates the situation in Continuous Spontaneous Localization theories \cite{Pearle:1988uh,Ghirardi:1989cn}.
(Energy conservation is explicitly violated in spontaneous-collapse theories such as the GRW model \cite{Ghirardi:1985mt}, and indeed searching for such violations is a known strategy for experimentally constraining such models \cite{vinante2016upper}.)
Aharonov, Popescu and Rohrlich  \cite{2016arXiv160905041A} argue that conservation laws can seemingly be violated in quantum mechanics, and label this a paradox.
Gao \cite{Gao:2016cmq} suggests a modified theory in which energy conservation is imposed in order to solve the preferred-basis problem.
Gisin and Cruzeiro \cite{gisin2018quantum} consider energy non-conservation in measurement of one spin in a long chain, but argue that conservation can be restored if an appropriate quantum clock is included.
Maudlin, Okon and Sudarsky \cite{Maudlin:2019bje} examine the definition of energy in a variety of contexts, focusing on semiclassical gravity, and conclude that there is ``no satisfactory way to define a (useful) notion of energy that is generically conserved.''
So{\l}tan et al. \cite{2019arXiv190706354S} argued for apparent nonconservation of energy by considering weak measurements.

In this paper we examine this issue in the context of standard, non-gravitational quantum theory based on Schr\"odinger evolution, and argue that observable energy non-conservation is a robust feature of quantum mechanics that might be experimentally accessible.
First, we analyze energy conservation during quantum measurement in the context of a well-defined formulation of quantum mechanics, specifically the Everett (Many-Worlds) interpretation.
We show that observers can indeed measure energy non-conservation, even though energy is conserved in the global wave function; as time passes, energy can be distributed differently among different branches.
Although the Everett approach allows us to analyze and reject the possibility that energy is simply lost in the apparatus or environment, the conclusion should apply to other formulations.
Second, we suggest a simple experimental protocol that demonstrates energy non-conservation in a controlled way, using a dipole-dipole interaction to entangle two spins and then measuring the spin of one, inducing a change in the energy of the other.
While our setup is idealized, we suggest that realistic experiments along these lines should be feasible.
Finally, we discuss some implications, including why energy conservation appears to be approximately true experimentally -- essentially because we tend not to come across, or unintentionally create, quantum states of the form (\ref{energysuperposition}) with wildly different values of the energy.
We also comment on the implications of this work for cosmological dynamics.

\section{An Everettian Example}

We start by considering energy conservation in Everettian quantum mechanics (EQM), or the Many-Worlds Interpretation \cite{Everett:1957hd,Wallace:2012zla}.
While our final results hold equally well for the Copenhagen formulation, EQM provides a completely-defined dynamical framework in which it is possible to rigorously follow the flow of energy between different parts of the wave function.

EQM is defined by positing that the state vector $|\psi\rangle$ is a complete representation of reality (i.e. it is a realist formulation with no hidden variables), and that the state always obeys the Schr\"odinger equation (i.e. there is no true wave function collapse, either spontaneous or induced),
\be
  \ham |\psi(t)\rangle = i \partial_t |\psi(t)\rangle.
\ee
In this approach, ``measurement'' occurs whenever a quantum system in superposition becomes entangled with its environment, leading to decoherence and branching of the wave function.

Let us consider a minimal explicit model of decoherence and branching, simple enough to analyze exactly but rich enough to include all of the relevant parts of the quantum state.
We factorize Hilbert space into a tensor product of system and environment, and consider a two-state system and a three-state environment, 
\begin{align}
 \hst &= \hst_s \otimes \hst_e, \nonumber\\
 \hst_s &= \mathrm{span}\{|1\rangle_s, |2\rangle_s\},\\
 \hst_e &= \mathrm{span}\{|0\rangle_e, |1\rangle_e, |2\rangle_e\}. \nonumber
\end{align}
The state starts with the system in a superposition and initially unentangled with the environment,
\be
  |\psi(0)\rangle = (\alpha|1\rangle_s +\beta |2\rangle_s)|0\rangle_e.
  \label{psi0}
\ee
Given appropriate interactions, under unitary evolution this evolves into
\be
  |\psi(1)\rangle = \alpha|1\rangle_s|1\rangle_e +\beta |2\rangle_s|2\rangle_e.
  \label{psi1}
\ee
If the environment states are orthogonal, $\langle1|2\rangle_e = 0$, decoherence has occurred and the state has branched into two distinct worlds.
(Realistic decoherence is more complicated and subtle, but this captures the basic idea.)

Given our definition of energy as the expectation value of the Hamiltonian (\ref{energydef}), from this discussion it should not be surprising that energy won't be conserved through the measurement process.
The energy in the global wave function $|\psi\rangle$ is constant over time, but after decoherence and branching the observers are either in the state $|1\rangle_s|1\rangle_2 $ or $|2\rangle_s|2\rangle_e$, each of which can have different energies.
Branching takes a fixed energy and distributes it unevenly between worlds.

In most discussions of quantum measurement this behavior is obscured, either because the system being measured is treated as an open system interacting with the outside world, or because the interaction Hamiltonian is modeled as time-dependent (in either of which cases, nobody should expect energy to be conserved).
In terms of the current example, we might imagine that there is some principle whereby differences in the energies of the environment states $\{\ket{i}_e\}$ precisely compensate for energy differences in the system states.
We can check that there is no such principle, and that environmental effects do not generally restore energy conservation, by constructing a simple but explicit model of a closed system with a complete time-independent Hamiltonian.

Consider the same Hilbert space as in the above example; now we will simply fill in what is meant by ``appropriate interactions.''
We imagine that the system basis states are energy eigenstates with different eigenvalues, corresponding to a ``self'' Hamiltonian
\be
\ham_\mathrm{self} = (E_1 |1\rangle_s\langle 1| + E_2 |2\rangle_s\langle 2|)\otimes \mathbb{1}_e.
\ee
This simply assigns energies $E_i$ to the states $|i\rangle_s$.
The interaction leaves the system states unchanged, but entangles them with different environment states. 
A simple choice is
\be
  \ham_\mathrm{int} = -i\lambda\Big[|1\rangle_s\langle 1|\otimes\big(|0\rangle_e\langle1| - |1\rangle_e\langle0|\big) + |2\rangle_s\langle 2|\otimes\big(|0\rangle_e\langle2| - |2\rangle_e\langle0|\big)\Big].
\ee
If the environment starts in the ready state $|0\rangle_e$, this will cause each system state $|i\rangle_s$ to become entangled with an environment state $|i\rangle_e$.
In this simplified context we leave out any discussion of a macroscopic measuring apparatus, pointer states, etc.

The sum $\ham = \ham_\mathrm{self} + \ham_\mathrm{int}$ can be exponentiated to obtain the unitary time-evolution operator,
\be
  \widehat{U}(t) = \exp(-i\ham t) = e^{-iE_1t} |1\rangle_s\langle1| \otimes
  \begin{pmatrix}
  \cos\lambda t & -\sin\lambda t & 0 \\ \sin\lambda t & \cos\lambda t & 0 \\ 0 & 0 & 1
  \end{pmatrix}
 + e^{-iE_2t} |2\rangle_s\langle2| \otimes
  \begin{pmatrix}
  \cos\lambda t &0& -\sin\lambda t  \\ 0 & 1 & 0 \\ \sin\lambda t &0& \cos\lambda t 
  \end{pmatrix}.
\ee
In this expression, the matrices act on the three-dimensional environment Hilbert space.
Starting from the initial state (\ref{psi0}), we obtain
\be
  |\psi(t)\rangle = \alpha e^{-iE_1t}|1\rangle_s(\cos \lambda t |0\rangle_e + \sin\lambda t|1\rangle_e)
  + \beta e^{-iE_2t}|2\rangle_s(\cos \lambda t |0\rangle_e + \sin\lambda t|2\rangle_e).
\ee
At time $t_*=\frac{\pi}{2\lambda}$, this becomes
\be
    |\psi(t_*)\rangle = \alpha e^{-iE_1t_*}|1\rangle_s|1\rangle_2 +\beta e^{-iE_2t_*}|2\rangle_s|2\rangle_e.
\ee
Up to phase factors, this matches the evolved state (\ref{psi1}).
Given the orthogonal environment states, we can consider this wave function as describing two decohered branches.
(In this simple setup, the two branches will begin to re-cohere after $t_*$.
It would be straightforward to complicate the model by putting more degrees of freedom into the environment, so that decoherence lasts an arbitrarily long time.)

We can now calculate the energies.
The expectation value of the interaction Hamiltonian vanishes,
\be
  \langle\psi |\ham_\mathrm{int}|\psi\rangle = 0,
\ee
leaving only the contribution from the self Hamiltonian,
\be
  \langle\psi |\ham_\mathrm{self}|\psi\rangle = |\alpha|^2 E_1 + |\beta|^2 E_2.
\ee
As expected, the energy changes from $|\alpha|^2 E_1 + |\beta|^2 E_2$ before branching to either $E_1$ or $E_2$ after.

This example demonstrates that there is no need for a compensating change in the energy of the environment or measuring apparatus when a system decoheres.
It also provides a satisfying explanation for the apparent violation of energy conservation: such violation is indeed only apparent, as the energy of the global wave function remains constant.
But observers don't see the global wave function, they only observe the branch they happen to be in.
And over the course of unitary evolution and branching, energy may be unevenly distributed among different branches.\footnote{The fact that energy is simply ``distributed among different branches'' depends on our assumption that the Hamiltonian included no off-diagonal matrix elements between different environment states. This is reasonable, but if it were not true, our result that observers can experience violations of energy conservation would be unaltered; we would merely lose the interpretation of the total energy as the sum of the energies of the branches, weighted by the amplitudes-squared.}

Of course EQM might not be the correct formulation of quantum mechanics, but a different theory shouldn't change the conclusion that observers see changes in the total energy of their universe.
In spontaneous-collapse models \cite{Ghirardi:1985mt,Pearle:1988uh,Ghirardi:1989cn}, energy is manifestly not conserved, even in the absence of interactions, and indeed this is a way of experimentally constraining such theories.
Hidden-variable models such as Bohmian mechanics \cite{duerr2009bohmian} are generally thought to have identical experimental predictions to Many-Worlds, but the situation is more subtle, because the wave function is not the only physical quantity.
It is therefore less clear how to define energy in such models \cite{Maudlin:2019bje}.
In epistemic theories \cite{leifer2014quantum} the wave function characterizes our ability to make predictions, rather than being a direct representation of physical reality; one might imagine an unknown mechanistic theory underlying some particular epistemic approach, but in the absence of specific models it is hard to make definitive statements.
In the standard textbook/Copenhagen approach, quantum systems are treated as open rather than closed, so there is no reason to expect energy to be conserved, as discussed.

\section{Toward a Realistic Experiment}

The next obvious question is whether we can experimentally observe violation of energy conservation.
The major challenge we face, especially for relatively small changes in the total energy, is the potential dissipation of energy in the process of measurement, which could be hard to keep track of precisely.
What we would like is for the process of measurement to be independent of the subsystem whose energy is changing.

This can be accomplished by entangling two systems 1 and 2 so that two different energy eigenstates of system 1 become entangled with two observably distinct states of system 2, then measuring system 2, collapsing system 1 into one or the other energy eigenstates.
We propose the following protocol:
\begin{enumerate}
\item Put a primary system 1 into a known quantum state that is a superposition of energy eigenstates.
\item Entangle that system with a probe system 2, in a way that does not involve substantial transfers of energy.
\item Measure the state of the probe system 2, again in a way that does not involve substantial transfers of energy.
\item Finish with the primary system 1 in an (at least approximate) energy eigenstate, with a substantially different value of the energy than the system started with.
\end{enumerate}
We will attempt to outline how such an experiment might be done.

Consider the following concrete example.
Take as the primary system a single spin-$\frac{1}{2}$ particle, such as a proton, with a nonzero charge and magnetic dipole moment $\mu_1$.
Keep the particle in a stationary quantum state, for example in a Penning trap.
Embed the particle in a magnetic field with magnitude $\vec B$ pointing in the $z$-direction.
For the probe system consider an uncharged spin-1/2 particle with a magnetic dipole moment $\mu_2$, such as a neutron.
The two particles experience a spin-spin interaction proportional to $\frac{1}{r^3}$, where $r$ is the relative distance between them.
This interaction causes the spins to become entangled.
We can then measure the spin of particle 2, causing particle 1 to collapse to a spin-$z$ eigenstate, so that its total energy changes.
In this section we do not explicitly include environmental degrees of freedom or calculate decoherence; because we can separate out the subsystem being observed from the one whose energy is not conserved, it is clear that the energy change is decoupled from the measurement process.

\begin{figure}[h]
\centering
\includegraphics[width=.6\textwidth]{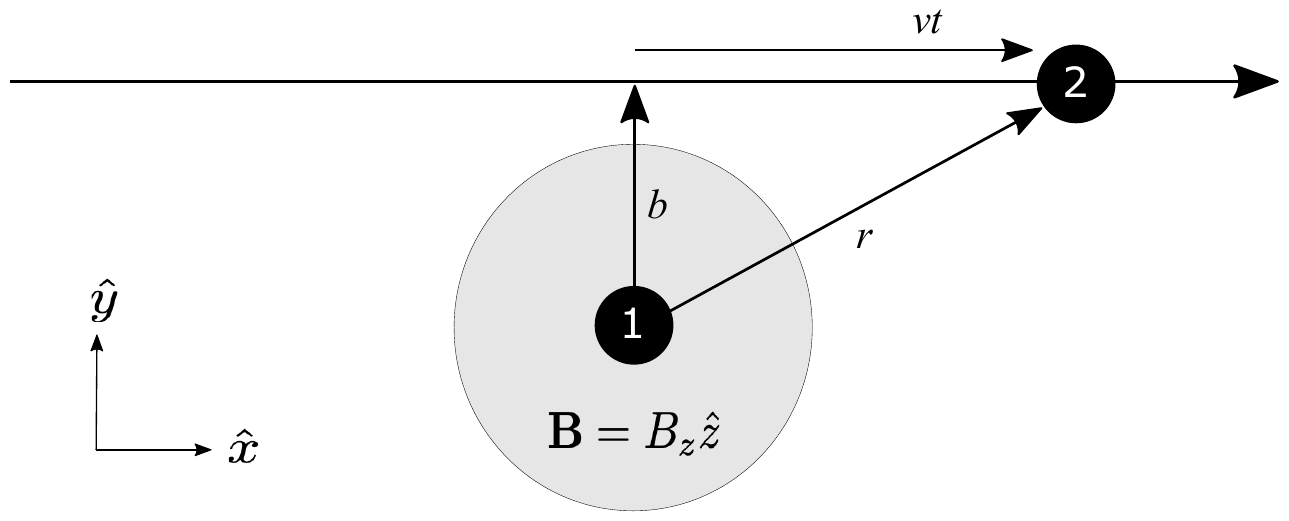}
\caption{Particle 1 is held trapped in a magnetic field in the $z$-direction, and put into a superposition of spin-$z$ (and thus energy) eigenstates.
Particle 2 passes by and becomes entangled via a spin-spin interaction. By appropriately measuring the spin of particle 2 once it has left the field, we can change the expectation value of the energy of particle 1, and thus of the system as a whole.}
\label{fig:experiment}
\end{figure}

We consider the spin-spin interaction between two magnetic dipoles.
The Hamiltonian we consider is therefore
\be
  \ham =  -({\vec\mu_1}\cdot\vec B) \otimes \mathbb{1}_2
  + \frac{g}{r^3}\left[\vec S_1 \cdot \vec S_2 - 3 (\vec S_1\cdot \hat{r})(\vec S_2 \cdot \hat{r})\right].
\ee
The first term is the interaction between particle 1 and the magnetic field, while the other terms describe an interaction between the two spins.
Here $\vec \mu_1$ is the magnetic moment for particle 1, $\vec B$ is the magnetic field value at $\vec r = 0$ (the location of particle 1),  $g$ is a coupling constant, and $r$ is the distance between the particles.
We imagine that the magnetic field points solely along the $z$-direction, $\vec B = B_z\hat{e}_z$.
The magnetic moment can be written in terms of the gyromagnetic ratio $\gamma_1$ as $\vec\mu_1 = \gamma_1\vec{S}$, where the spin operator for a spin-1/2 particle is $\vec{S} = \frac{1}{2}\vec\sigma$.
Then we can define the Larmor frequency as $\omega = -\gamma_1 B_z$.
The Hamiltonian can therefore be written
\be
  \ham =  \omega S_1^z\otimes\mathbb{1}_2 
  + \frac{g}{r^3}\left[\vec S_1 \cdot \vec S_2 - 3 (\vec S_1\cdot \hat{r})(\vec S_2 \cdot \hat{r})\right].
  \label{hamiltonian}
\ee
For simplicity we imagine that particle 2 never enters the magnetic field, so that interaction is not included, though it would be straightforward to do so.
We also do not include the energy of the $B$-field, as that is taken to be stationary and does not exchange energy with the particles.
(There will be a back-reaction of the magnetic moment on the magnetic field, but that is subdominant to the interaction energy we consider here.)
Two nucleons can of course interact via the nuclear force, which is much stronger than the spin-spin interaction at short distances, but that takes an exponentially-damped Yukawa form, so it can be neglected if the impact parameter is large on nuclear length scales.

We do not explicitly include the kinetic-energy terms for the individual particles, since we approximate their velocities as constant.
We assume that both particles are described by localized wave packets that can be approximated as single points as far as the interaction is concerned, rather than explicitly considering their spatial wave functions.
We imagine that particle 1 remains stationary in space, while particle 2 moves on a straight line in the $x$-direction with constant velocity $v$ and impact parameter $b$, so that $r=\sqrt{b^2 + v^2t^2}$.
This will not strictly be true, since the dipole-dipole interaction induces a slight force between the two particles.
It is nevertheless a useful approximation to make our point here, since the interaction potential between the particles is independent of the strength of the $B$-field, and it is the coupling of particle 1 to that field that will ultimately lead to a change in the total energy. 
For thought-experiment purposes, we can imagine cranking up the $B$-field so high that minor perturbations to the kinetic energies of the particles are irrelevant. Here we need only show that the interaction will entangle the two spins.

At the beginning and end of the experiment, the two particles are sufficiently separated that their interaction can be ignored.
Particle 1, trapped in the magnetic field, has energy eigenstates $\ket\uparrow_1 = \ket{+z}_1$ and $\ket\downarrow_1=\ket{-z}_1$, with energies 
\be
E_\pm = \pm \frac{1}{2}\omega.
\label{energy_eigenvalues}
\ee
Since the kinetic energy of particle 2 is assumed to be constant, particle 1 is our primary system whose energy will change.\footnote{It would be important to ensure that the spin states of particle 1 did not decohere when it was put into the magnetic field, in effect measuring the spin before the experiment started.
If that were the case, the particle would simply be taking different amounts of energy from the magnetic field in two distinct branches of the wave function.}

Let us write the time integral of the Hamiltonian as a matrix in the $\{\ket{\uparrow\uparrow}, \ket{\uparrow\downarrow}, \ket{\downarrow\uparrow}, \ket{\downarrow\downarrow}\}$ basis as
\be
   \int_{t_0}^t \ham(t') \, dt' = 
   \frac{1}{4}\left(
  \begin{array}{cccc}
  \theta+2\Omega &&& \xi^* \\
  & -\theta+2\Omega & -\theta & \\
  & -\theta & -\theta-2\Omega & \\
  \xi &&& \theta-2\Omega
  \end{array}
  \right).
\ee
Here we have defined 
\be
  \theta(t)=\int_{t_0}^t \lambda[r(t')] \, dt' = \frac{gt}{b^2\sqrt{b^2+v^2t^2}} + \frac{g}{b^2v},
  \label{theta}
\ee
\be
 \xi(t) = -3\int_{t_0}^t \frac{\lambda(t') (vt' + ib)^2}{v^2t'^2+ b^2}  \, dt' = \frac{g (b-i t v) (2 i b-tv)}{b^2 v (b+i t v) \sqrt{b^2+t^2 v^2}} + \frac{g}{b^2 v},
 \label{xi}
\ee
and 
\be
   \Omega(t) = \omega(t-t_0).
\ee
On the right-hand sides, we have assumed that particle 2 starts sufficiently far away that the $t_0$ boundary terms are negligible (i.e. we can set $t_0$ to $-\infty$).
Note that the apparent time-dependence of $\ham$ merely reflects the fact that the particles are moving with respect to each other; no energy is being put into or taken from the system.

The spin states of both particles are initially taken to be equal superpositions of $\ket{\uparrow}\equiv\ket{+z}$ and $\ket{\downarrow} \equiv \ket{-z}$.
Particle 1 is embedded in a magnetic field pointing in the $z$-direction, and will therefore be precessing.
Since we might not know the state of its precession, we allow for an arbitrary phase $\phi_0$ for particle 1.
The initially unentangled state is therefore
\bea
  \ket{\psi_0} &=& \frac{1}{2}\left(\ket{\uparrow} + e^{i\phi_0} \ket{\downarrow}\right)_1\left(\ket{\uparrow} + \ket{\downarrow}\right)_2 \\
  &=&  \frac{1}{2}\left(\ket{\uparrow\uparrow} + \ket{\uparrow\downarrow} + e^{i\phi_0} \ket{\downarrow\uparrow} + e^{i\phi_0} \ket{\downarrow\downarrow}\right),
\eea
so that both spins start out unentangled and perpendicular to $\ket{+z}$.

We can calculate the time-evolution operator 
\be
\widehat{U}(t) = \exp\left[-i\int \ham(t') \, dt'\right], 
\ee
but we won't bother writing the matrix elements explicitly.
For convenience we write the wave function as
\be
   \ket{\psi(t)}= \psiuu \ket{\uparrow\uparrow} + \psiud\ket{\uparrow\downarrow} + \psidu\ket{\downarrow\uparrow} + \psidd\ket{\downarrow\downarrow} .
\ee
The components of $\ket{\psi(t)} = \widehat{U}(t)\ket{\psi_0}$ are
\bea
  \psiuu &=& \frac{1}{8}e^{-i\theta/4}\left[4\cos{\Delta_\xi}
      - i\left(\frac{\sin\Delta_\xi}{\Delta_\xi}\right)
      \left(2\Omega + e^{i\phi_0} \xi^*\right) 
    \right] , \nonumber\\
  \psiud &=& \frac{1}{8}e^{i\theta/4}\left[4\cos{\Delta_\theta}
      - i\left(\frac{\sin\Delta_\theta}{\Delta_\theta}\right)
      \left(2\Omega - e^{i\phi_0} \theta\right) 
   \right],  \nonumber\\
  \psidu &=&  \frac{1}{8}e^{i\theta/4}\left[4e^{i\phi_0}\cos{\Delta_\theta}
      + i\left(\frac{\sin\Delta_\theta}{\Delta_\theta}\right)
      \left(2e^{i\phi_0}\Omega +  \theta\right) 
   \right] , \nonumber\\
  \psidd &=& \frac{1}{8}e^{-i\theta/4}\left[4e^{i\phi_0}\cos{\Delta_\xi}
      + i\left(\frac{\sin\Delta_\xi}{\Delta_\xi}\right)
      \left(2 e^{i\phi_0}\Omega - \xi\right) 
   \right],
   \label{psioft}
\eea
where we have defined
\be  
  \Delta_\xi = \frac{1}{4}\sqrt{|\xi|^2 + 4\Omega^2},
  \qquad
  \Delta_\theta = \frac{1}{4}\sqrt{\theta^2 + 4\Omega^2}.
\ee

The initial energy of particle 1 is 
\be
  E_1(t_0) = 0.
\ee
Because it is outside the magnetic field, and we approximate its kinetic energy as constant during the entire experiment, the total energy of particle 2 is the same at initial and final times,
\be
  E_2(t_0) = E_2(t_f).
\ee
Within these approximations, then, the change in total energy after measurement will be the change in the energy of the stationary particle 1.

The reduced density matrix $\hat\rho_1 = \Tr_2\ket{\psi}\bra{\psi}$ for particle 1 is
\be
   \hat\rho_1= \left(
  \begin{array}{cc}
  |\psiuu|^2 + |\psiud|^2 & \psiuu\psidu^* + \psiud\psidd^* \\
  \psiuu^*\psidu + \psiud^*\psidd & |\psidu|^2 + |\psidd|^2 
  \end{array}
  \right),
\ee
with eigenvalues
\be
  k_\pm = \frac{1}{2}\left( 1 \pm \sqrt{1 - 4|\psiuu\psidd - \psiud\psidu|^2}\right).
\ee
The two spins will be maximally entangled when both eigenvalues are 1/2. 
This will permit us to indirectly measure the spin of particle 1 by measuring that of particle 2.

In the late-time $t\rightarrow +\infty$ limit we have
\bea
  \theta_\infty = \xi_\infty &=& \frac{2g}{b^2v}, \\
  \Delta_{\theta,\xi}(t\rightarrow \infty) &=& \frac{1}{2}\omega  t.
\eea
The components of the spin wave function then simplify to
\bea
  \psiuu(\infty) &=& \frac{1}{2}e^{-i\theta_\infty/4}  e^{-i\omega t/2}, \nonumber\\
  \psiud(\infty) &=& \frac{1}{2}e^{i\theta_\infty/4}  e^{-i\omega t/2} , \nonumber\\
  \psidu(\infty) &=& \frac{1}{2}e^{i\theta_\infty/4} e^{i\phi_0} e^{i\omega t/2}, \nonumber\\
  \psidd(\infty) &=& \frac{1}{2}e^{-i\theta_\infty/4} e^{i\phi_0} e^{i\omega t/2},
   \label{psiinfinity}
\eea
and the eigenvalues are
\be
  k_\infty = \frac{1}{2}\left(1\pm \cos\frac{\theta_\infty}{2}\right).
\ee
The final eigenvalues are independent of the initial phase $\phi_0$ of the particle in the magnetic field.
Maximal entanglement therefore occurs whenever $\theta_\infty = (2n+1)\pi$ for any integer $n$.
This can be ensured by careful choice of the impact parameter and velocity, since $\theta_\infty = 2g/b^2v$.

If we choose parameters such that $\theta_\infty=\pi$, the final state can be written
\be
  \ket{\psi_\infty} = \frac{1}{\sqrt{2}}\left[e^{-i(\pi/4+\omega t/2)}\ket{\uparrow}_1\ket{+y}_2
  + e^{i(\pi/4+\phi_0+\omega t/2)}\ket{\downarrow}_1\ket{-y}_2\right].
\ee
where $\ket{\pm y} = (\ket{\uparrow}\pm i \ket{\downarrow})/\sqrt{2}$.
The $z$-eigenstates of particle 1 are thus entangled with $y$-eigenstates of particle 2.

Now we can measure the spin of particle 2 along the $y$-axis, well after it has moved away from particle 1.
The measurement process may transfer energy from particle 2 to the environment, but the amount of energy should be independent of the measurement outcome.
If we measure particle 2 to be spin-up, we know that particle 1 will be in a spin-up eigenstate in the direction $z$ of the magnetic field, and likewise for spin-down.
Particle 1 is therefore in an energy eigenstate with energy (\ref{energy_eigenvalues}), while particle 2 has spin energy zero (and the same kinetic energy it started with).
The total energy of the system (and therefore the universe) will have shifted by an amount
\be
  \Delta E = \Delta \langle\widehat H\rangle = \pm \frac{1}{2}\gamma_1 B_z,
\ee
violating conservation of total energy in our observable universe.

A realistic experiment along these lines seems challenging, as the dipole-dipole interaction is weak, and we have made various approximations along the way.
For a realistic value of the coupling constant $g= 0.557\,$GeV$^{-2}$ \cite{levitt2015, gamman, gammap}, the maximal-entanglement condition on $\theta_\infty$ implies that the impact parameter and velocity should satisfy $b\sqrt{v} = 7.38\times 10^{-14}\,$cm, where $v$ is measured with respect to the speed of light. 
This distance scale is smaller than the size of a proton; to have a larger impact parameter, and therefore avoid having the spin-spin interaction be swamped by nuclear forces, we would have to choose $v$ to be very small.
Controlling the relevant experimental parameters to the necessary precision would seem challenging, to put it lightly.

We are nevertheless encouraged that a protocol analogous to the one presented here is possible in principle, and the fact that the final answer scales with $B_z$, while the quantities we neglected in our approximations are independent of this quantity, suggests that the violation of energy conservation can plausibly be made large enough to dominate over any errors introduced by measurement procedures or approximation schemes.

\section{Discussion}

Conservation of energy is a cherished principle in physics, but it does have its limitations.
Here we have explored such a limitation in the context of quantum measurement, a process which (at least from the perspective of a particular observer) is non-unitary and irreversible, not to mention incompletely understood.
We have established that any hope that energy conservation might be preserved by including the environment and measuring apparatus will not pan out, and suggested an experimental protocol by which this violation might be measured.
There might nevertheless be a sense in which energy is conserved: the total energy of the universal wave function in Everettian quantum mechanics is constant as long as the Hamiltonian is independent of time, even as that energy is distributed differently through branches of the wave function over time.
(This might be taken as an argument in favor of the Everett formulation, if one thinks there should be a simple definition of energy that is unambiguously conserved in the universe as a whole.)

It might be asked whether a similar analysis would apply to other conservation laws, such as conservation of electric charge.
But in this case there is an important difference: we expect the universe to be in an exact eigenstate of charge (presumably with zero total charge), rather than a superposition of different eigenstates.
In an Everettian picture, every branch of the universe would therefore have the same charge.
By contrast, it is important that our universe not be in an eigenstate of energy, otherwise there would be no nontrivial time evolution.
(Quantum gravity governed by the Wheeler-DeWitt equation, $\ham\Psi = 0$, is a subtle case \cite{Halliwell:2002th}. 
For some relevant debate see \cite{Boddy:2014eba,Lloyd:2016ahu}. 
One approach is to recover an emergent time parameter $\tau$ by writing $\ham = \ham_\mathrm{eff} - i {d}/{d\tau}$; our analysis would then apply to energy defined by the effective Hamiltonian $\ham_\mathrm{eff}$.)

A useful way of thinking about what our result means for the evolution of the wave function of the universe is to work in the energy eigenbasis of the total Hamiltonian.
The wave function as a whole is some superposition spanning a subset of the entire basis.
In the Everett picture, as time passes, the overall state splits into branches.
Each branch will be a superposition of some of the original energy eigenstates, but some eigenstates will have much lower amplitudes, and even drop out of the superposition entirely if the amplitude goes to zero.
Branching therefore filters the original set of eigenstates in the wave function of the universe down to an ever-smaller set of eigenstates with nonzero support in any particular branch.

\begin{figure}[h]
\centering
\includegraphics[width=.6\textwidth]{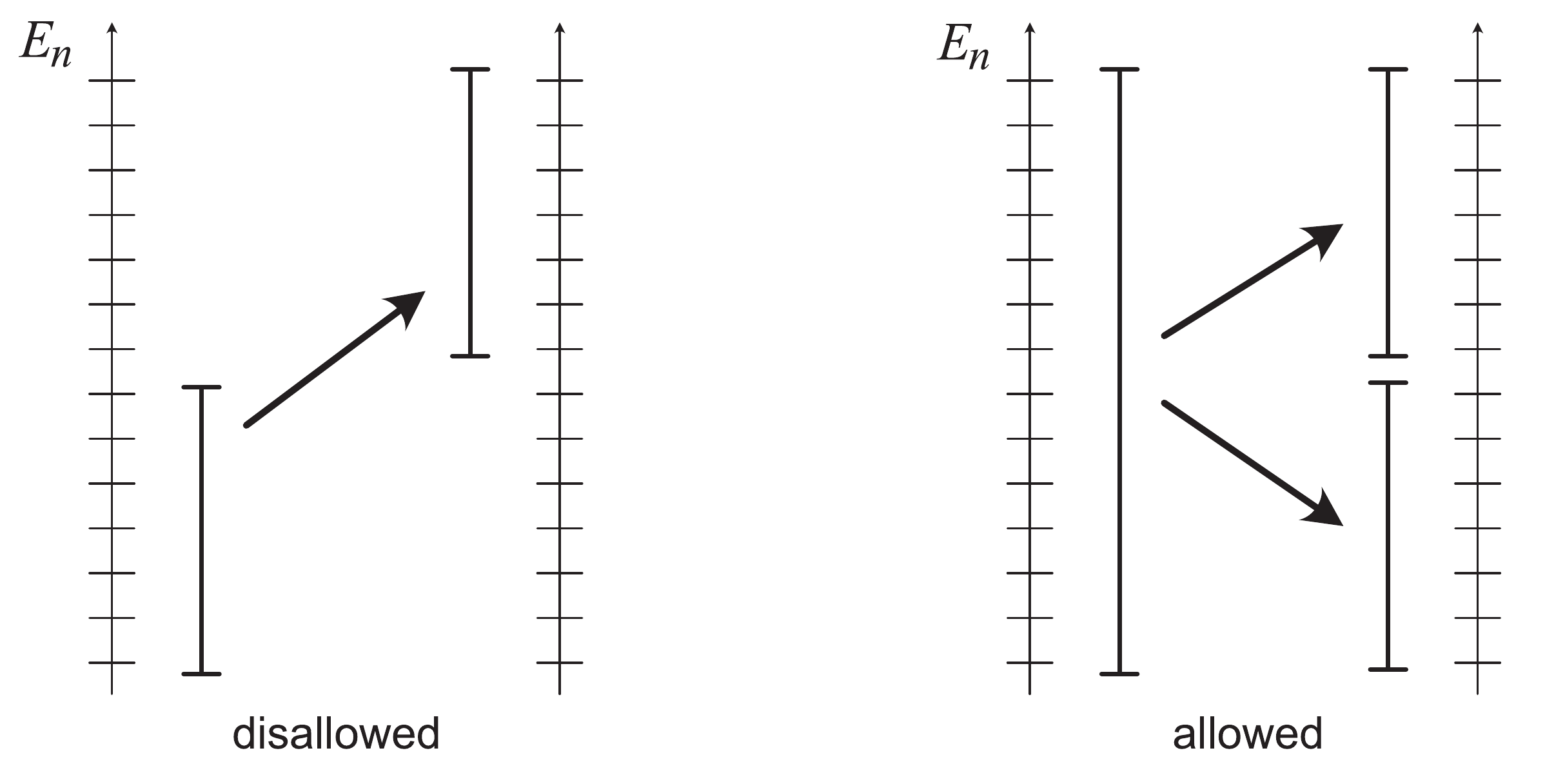}
\caption{A measurement of total energy at one moment in time cannot disagree with a measurement at a later time \cite{griffiths,hartle1995conservation}.
This is compatible with the kind of evolution considered in this paper, in which branching takes a superposition of some energy eigenstates to multiple different superpositions, with different expectation values of the energy in each.
The figure shows this schematically: first a disallowed evolution from one superposition of energies to a distinct superposition, then an allowed branching from one superposition to others formed from different subsets of the original eigenstates.}
\label{fig:allowed}
\end{figure}

This viewpoint illustrates how our claim of energy non-conservation is compatible with the argument for conservation in \cite{griffiths,hartle1995conservation}.
What that argument rules out is a fluctuation from one energy value to another; in particular, histories defined by projections onto incompatible values of the total energy will not decohere.
This is compatible with our result, as illustrated schematically in Figure~\ref{fig:allowed}.
The expectation value of the Hamiltonian can still change within histories that do decohere, for example when one world branches into two worlds constructed from different subsets of the original energy eigenstates.

Our scenario does not give any hope for a perpetual motion machine.
Violations of energy conservation are subject to quantum randomness, so that the total energy of the universe would be expected to increase or decrease with roughly equal probability, in an unpredictable way.\footnote{Decoherence and the linearity of the Schr\"odinger equation ensure that communication between different branches of the wave function is not possible. If it were, we might imagine systematically transferring energy from one branch to another -- an ``Everett pipeline,'' analogous to Polchinski's ``Everett phone'' \cite{Polchinski:1990py}. This might be an argument in favor of the linearity of quantum mechanics, if one wanted to avoid perpetual motion machines. We thank Grant Remmen for this observation.}
The size of such fluctuations will tend to decrease over time, as decoherence leads to approximately-classical looking states constructed from eigenstates of nearly-equal energies.

This helps explain why energy seems to be conserved to a high degree of accuracy in experiments.
The universe evolves into a set of branches, each of which is a superposition of eigenstates with total energies that are extremely close to each other, and thus to the classical notion of a conserved ``energy of the universe.''
According to our suggested experimental protocol, large violations of energy conservation would only happen when we observe initially unentangled quantum systems that are superpositions of very different energies.
In practice this is hard to achieve, since macroscopic systems tend to decohere very quickly.

The idea that decoherence produces branches of the wave function with approximately well-defined energies also suggests a dynamical selection process for Everettian worlds.
Even if we start with a wave function that involves a superposition of states with different energies, branches where the energies are very different will rapidly decohere from each other.
Remaining branches will be constructed from eigenstates of the Hamiltonian with approximately similar energy eigenvalues.
This implies that our Hamiltonian -- the laws of physics of our observed world -- might represent a subset of the total Hamiltonian of the universe.
It will be interesting to examine the ramifications of this idea for the project of seeing how spacetime and quantum fields can emerge from the spectrum of the Hamiltonian \cite{Cao:2016mst,Cao:2017hrv,Cotler:2017abq,Carroll:2018rhc,Carroll:2021aiq}.

\section*{Acknowledgements}
We would like to thank Anthony Bartolotta, Brad Fillipone, Jason Pollack, Ken Olum, Jonathan Oppenheim, Gil Refael, Grant Remmen, Ashmeet Singh, and Mark Wise for helpful discussions during the course of this project. This research is funded in part by the Walter Burke Institute for Theoretical Physics at Caltech, by the U.S. Department of Energy, Office of Science, Office of High Energy Physics, under Award Number DE-SC0011632, and by the Foundational Questions Institute.

\bibliographystyle{utphys}
\bibliography{energyconservation}

\providecommand{\href}[2]{#2}\begingroup\raggedright\begin{thebibliography}{10}

\bibitem{Wallace2007-WALTQM}
D.~Wallace, ``The quantum measurement problem: State of play,'' in {\em The
  Ashgate Companion to Contemporary Philosophy of Physics}, D.~Rickles, ed.
\newblock Ashgate, 2007.

\bibitem{hartle1995conservation}
J.~B. Hartle, R.~Laflamme, and D.~Marolf, ``Conservation laws in the quantum
  mechanics of closed systems,'' {\em Physical Review D} {\bfseries 51} no.~12,
  (1995) 7007, \href{http://arxiv.org/abs/gr-qc/9410006}{{\ttfamily
  arXiv:gr-qc/9410006 [gr-qc]}}.

\bibitem{griffiths}
R.~Griffiths. Unpublished.

\bibitem{Pearle:2000qb}
P.~M. Pearle, ``{Wave function collapse and conservation laws},''
  \href{http://dx.doi.org/10.1023/A:1003677103804}{{\em Found. Phys.}
  {\bfseries 30} (2000) 1145--1160},
\href{http://arxiv.org/abs/quant-ph/0004067}{{\ttfamily arXiv:quant-ph/0004067
  [quant-ph]}}.

\bibitem{Pearle:1988uh}
P.~M. Pearle, ``{Combining Stochastic Dynamical State Vector Reduction With
  Spontaneous Localization},''
  \href{http://dx.doi.org/10.1103/PhysRevA.39.2277}{{\em Phys. Rev. A}
  {\bfseries 39} (1989) 2277--2289}.

\bibitem{Ghirardi:1989cn}
G.~C. Ghirardi, P.~M. Pearle, and A.~Rimini, ``{Markov Processes in Hilbert
  Space and Continuous Spontaneous Localization of Systems of Identical
  Particles},'' \href{http://dx.doi.org/10.1103/PhysRevA.42.78}{{\em Phys. Rev.
  A} {\bfseries 42} (1990) 78--79}.

\bibitem{Ghirardi:1985mt}
G.~Ghirardi, A.~Rimini, and T.~Weber, ``{A Unified Dynamics for Micro and MACRO
  Systems},'' \href{http://dx.doi.org/10.1103/PhysRevD.34.470}{{\em Phys. Rev.
  D} {\bfseries 34} (1986) 470}.

\bibitem{vinante2016upper}
A.~Vinante, M.~Bahrami, A.~Bassi, O.~Usenko, G.~Wijts, and T.~Oosterkamp,
  ``Upper bounds on spontaneous wave-function collapse models using
  millikelvin-cooled nanocantilevers,'' {\em Physical Review Letters}
  {\bfseries 116} no.~9, (2016) 090402,
  \href{http://arxiv.org/abs/1510.05791}{{\ttfamily arXiv:1510.05791
  [quant-ph]}}.

\bibitem{2016arXiv160905041A}
Y.~{Aharonov}, S.~{Popescu}, and D.~{Rohrlich}, ``{On conservation laws in
  quantum mechanics},'' {\em arXiv e-prints} (Sep, 2016) arXiv:1609.05041,
  \href{http://arxiv.org/abs/1609.05041}{{\ttfamily arXiv:1609.05041
  [quant-ph]}}.

\bibitem{Gao:2016cmq}
S.~Gao, \href{http://dx.doi.org/10.1017/9781316407479}{{\em {The Meaning of the
  Wave Function}}}.
\newblock Cambridge University Press, 2018.
\newblock \href{http://arxiv.org/abs/1611.02738}{{\ttfamily arXiv:1611.02738
  [quant-ph]}}.
\newblock
\url{http://www.cambridge.org/academic/subjects/physics/history-philosophy-and-foundations-physics/meaning-wave-function-search-ontology-quantum-mechanics?format=HB&isbn=9781107124356}.
\newblock

\bibitem{gisin2018quantum}
N.~Gisin and E.~Zambrini~Cruzeiro, ``Quantum measurements, energy conservation
  and quantum clocks,'' {\em Annalen der Physik} {\bfseries 530} no.~6, (2018)
  1700388.

\bibitem{Maudlin:2019bje}
T.~Maudlin, E.~Okon, and D.~Sudarsky, ``{On the Status of Conservation Laws in
  Physics: Implications for Semiclassical Gravity},''
\href{http://arxiv.org/abs/1910.06473}{{\ttfamily arXiv:1910.06473 [gr-qc]}}.

\bibitem{2019arXiv190706354S}
S.~{So{\l}tan}, M.~{Fr{\k{a}}czak}, W.~{Belzig}, and A.~{Bednorz}, ``{Is energy
  conserved when nobody looks?},'' {\em arXiv e-prints} (July, 2019)
  arXiv:1907.06354, \href{http://arxiv.org/abs/1907.06354}{{\ttfamily
  arXiv:1907.06354 [quant-ph]}}.

\bibitem{Everett:1957hd}
H.~Everett, ``{Relative state formulation of quantum mechanics},''
\href{http://dx.doi.org/10.1103/RevModPhys.29.454}{{\em Rev. Mod. Phys.}
  {\bfseries 29} (1957) 454--462}.

\bibitem{Wallace:2012zla}
D.~Wallace, {\em {The emergent multiverse: quantum theory according to the
  Everett Interpretation}}.
\newblock Oxford University Press, Oxford, 2012.
\newblock
\url{http://www-spires.fnal.gov/spires/find/books/www?cl=QC174.12.W348::2012}.
\newblock

\bibitem{duerr2009bohmian}
D.~Duerr, S.~Goldstein, R.~Tumulka, and N.~Zanghi, ``Bohmian mechanics,''
  \href{http://arxiv.org/abs/0903.2601}{{\ttfamily arXiv:0903.2601
  [quant-ph]}}.

\bibitem{leifer2014quantum}
M.~S. Leifer, ``Is the quantum state real? an extended review of
  $\psi$-ontology theorems,'' {\em Quanta} {\bfseries 3(1)} (2014) 67,
  \href{http://arxiv.org/abs/arXiv:1409.1570}{{\ttfamily arXiv:1409.1570}}.

\bibitem{levitt2015}
M.~H. Levitt, {\em Spin dynamics: basics of nuclear magnetic resonance}.
\newblock Wiley, 2015.

\bibitem{gamman}
``Codata value: neutron gyromagnetic ratio.''
\newblock \url{https://physics.nist.gov/cgi-bin/cuu/Value?gamman}.

\bibitem{gammap}
``Codata value: proton gyromagnetic ratio.''
\newblock \url{https://physics.nist.gov/cgi-bin/cuu/Value?gammap}.

\bibitem{Halliwell:2002th}
J.~Halliwell and J.~Thorwart, ``{Life in an energy eigenstate: Decoherent
  histories analysis of a model timeless universe},''
  \href{http://dx.doi.org/10.1103/PhysRevD.65.104009}{{\em Phys. Rev. D}
  {\bfseries 65} (2002) 104009},
  \href{http://arxiv.org/abs/gr-qc/0201070}{{\ttfamily arXiv:gr-qc/0201070}}.

\bibitem{Boddy:2014eba}
K.~K. Boddy, S.~M. Carroll, and J.~Pollack, ``{De Sitter Space Without
  Dynamical Quantum Fluctuations},''
  \href{http://dx.doi.org/10.1007/s10701-016-9996-8}{{\em Found. Phys.}
  {\bfseries 46} no.~6, (2016) 702--735},
  \href{http://arxiv.org/abs/1405.0298}{{\ttfamily arXiv:1405.0298 [hep-th]}}.

\bibitem{Lloyd:2016ahu}
S.~Lloyd, ``{Decoherent histories approach to the cosmological measure
  problem},'' \href{http://arxiv.org/abs/1608.05672}{{\ttfamily
  arXiv:1608.05672 [quant-ph]}}.

\bibitem{Polchinski:1990py}
J.~Polchinski, ``{Weinberg's nonlinear quantum mechanics and the EPR
  paradox},''
\href{http://dx.doi.org/10.1103/PhysRevLett.66.397}{{\em Phys. Rev. Lett.}
  {\bfseries 66} (1991) 397--400}.

\bibitem{Cao:2016mst}
C.~Cao, S.~M. Carroll, and S.~Michalakis, ``{Space from Hilbert Space:
  Recovering Geometry from Bulk Entanglement},''
  \href{http://dx.doi.org/10.1103/PhysRevD.95.024031}{{\em Phys. Rev.}
  {\bfseries D95} no.~2, (2017) 024031},
\href{http://arxiv.org/abs/1606.08444}{{\ttfamily arXiv:1606.08444 [hep-th]}}.

\bibitem{Cao:2017hrv}
C.~Cao and S.~M. Carroll, ``{Bulk entanglement gravity without a boundary:
  Towards finding Einstein\textquoteright{}s equation in Hilbert space},''
  \href{http://dx.doi.org/10.1103/PhysRevD.97.086003}{{\em Phys. Rev. D}
  {\bfseries 97} no.~8, (2018) 086003},
  \href{http://arxiv.org/abs/1712.02803}{{\ttfamily arXiv:1712.02803
  [hep-th]}}.

\bibitem{Cotler:2017abq}
J.~S. Cotler, G.~R. Penington, and D.~H. Ranard, ``{Locality from the
  Spectrum},'' \href{http://dx.doi.org/10.1007/s00220-019-03376-w}{{\em Commun.
  Math. Phys.} {\bfseries 368} no.~3, (2019) 1267--1296},
  \href{http://arxiv.org/abs/1702.06142}{{\ttfamily arXiv:1702.06142
  [quant-ph]}}.

\bibitem{Carroll:2018rhc}
S.~M. Carroll and A.~Singh, ``{Mad-Dog Everettianism: Quantum Mechanics at Its
  Most Minimal},'' in {\em What Is Fundamental?}, A.~Aguirre, B.~Foster, and
  Z.~Merali, eds., pp.~95--104.
\newblock Springer, 2019.
\newblock \href{http://arxiv.org/abs/1801.08132}{{\ttfamily arXiv:1801.08132
  [quant-ph]}}.

\bibitem{Carroll:2021aiq}
S.~M. Carroll, ``{Reality as a Vector in Hilbert Space},''
  \href{http://arxiv.org/abs/2103.09780}{{\ttfamily arXiv:2103.09780
  [quant-ph]}}.

\end{thebibliography}\endgroup

\end{document}